\documentclass[12pt, draftclsnofoot, onecolumn]{IEEEtran}
\usepackage{cite}
\usepackage{microtype}
\usepackage{amsmath}
\usepackage{accents}
\usepackage{amssymb}
\usepackage{graphicx}
\usepackage{suffix}
\usepackage{mathtools}
\usepackage{stackengine}
\usepackage{caption}
\usepackage{subcaption}
\usepackage[export]{adjustbox}

\newcommand{\meijerG}[7]{\text G^{#1,#2}_{#3,#4} \left. \Big[ #7 \middle\vert \right.  \begin{smallmatrix} #5 \\ #6 \end{smallmatrix} \Big]}

\newcommand{\addtag}{\refstepcounter{equation}\tag{\theequation}}

\usepackage{scalerel,stackengine}

\begin{document}
	\title{\huge{{Impact of RF I/Q Imbalance on Interference-Limited Mixed RF/FSO TWR Systems with Non-Zero Boresight Error}}}
	\iffalse	
	\author{Author 1, \and Author 2, Author 3 and Author 4 \thanks{} \thanks{ .}	
	} \fi

\author{Abhijeet Upadhya,  \and Juhi Gupta, \textit{Member, IEEE}, \and Vivek K. Dwivedi, \textit{Member, IEEE,} \and\\  and Mohamed-Slim Alouini,\textit{ Fellow, IEEE}  \thanks{Abhijeet Upadhya is with department of Electronics and Communication Engineering, Ajay Kumar Garg Engineering College, Ghaziabad, India, (Email: upadhya.abhijeet@gmail.com)} \thanks{Juhi Gupta and Vivek K. Dwivedi are with department of Electronics and Communication Engineering, Jaypee Institute of Information Technology, Noida, India, (Email: juhi@jiit.ac.in, vivek.dwivedi@jiit.ac.in).} \thanks{Mohamed-Slim Alouini is with Computer, Electrical, and Mathematical Sciences and Engineering Division at King Abdullah University of Science and Technology (KAUST), Thuwal, Makkah Province, Saudi Arabia (email: slim.alouini@kaust.edu.sa).}}

	\maketitle
	
\begin{abstract}
	In this letter, we investigate a generic model assessing the effect of in-phase/quadrature-phase imbalance (IQI) on an asymmetric dual hop radio frequency/free space optical (RF/FSO) two-way relay (TWR) system in the presence of multiple co-channel interferers (CCIs) at the relay. The fading on the RF and FSO links have been modeled using \(K\)-distribution and double generalized Gamma (D-GG) turbulence model, respectively. The impact of non-zero boresight pointing error and type of optical demodulation schemes have been considered on the FSO link. To this end, a closed-form probability density function (PDF) has been derived for the FSO link undergoing D-GG irradiance with non-zero boresight pointing error. Furthermore, the exact and high signal-to-noise (SNR) asymptotic expression for outage probability have been presented. To gain insights into the throughput of the system, approximate and asymptotic expressions for the achievable sum rate (ASR) have been derived. The results show dependency of reliability and throughput offered on IQI at the RF front-end and strength of interference for the considered TWR system.
\end{abstract}
\section{Introduction}
\IEEEPARstart{I}{n} order to improve data rates for wireless users, the most power-efficient and reliable mode is to introduce cooperative communications. An asymmetric cooperative RF/FSO (radio frequency/free space optics) relaying network provides efficient way to achieve last-mile connectivity and is an effective solution to overcome the connectivity gap between RF and FSO backbone networks. In the context of cooperative relaying, two way relaying (TWR) approach has the capability to improve the spectral efficiency compared to one way relaying (OWR) \cite{E}. In most of previous works in mixed RF/FSO relaying systems including \cite{DGGwithPointError, mypaper4, Aloiuni-3-QAM, Aloiuni-1-phyc-sec}, the RF link hardware has been assumed to perform ideally. However, due to cost constraints and limited accuracy of the analog hardware, in practice, RF impairments such as flicker noise, in-phase (I) and quadrature-phase (Q) imbalances (IQI), current/voltage offsets due to self-mixing, phase noise, and power amplifier nonlinearities impact the overall performance of relaying systems \cite{F, Aloiuni-4-HPA, M}. \par 
The mismatch between I and Q components in the transmitter (Tx) and receiver (Rx) cause IQI which is the most detrimental among all possible analog hardware impairments. Ideally, the I and Q components are equal in amplitude and have \(90^{\circ}\) phase difference. However, due to high switching speeds, up- and down-converters involved in RF front-end offer non-zero IQI in practical circuits \cite{G}. Additionally, practical mixed RF/FSO relaying systems are affected by the presence of interference from neighboring communication systems \cite{mypaper3}. 
Besides, apart from atmospheric turbulence, the FSO link is affected by the pointing error which occurs due to boresight and jitter that originate due to the sway, vibration, and thermal expansion of buildings \cite{A, Aloiuni-2-PRS}. However, the non-zero boresight error occurs due to the fixed displacement between centers of beam and detector and is a quite realistic consideration in the pointing error model \cite{A, Aloiuni-2-PRS, mypaper4}, where the effect of different jitters for the elevation and the horizontal displacements are assumed.   \par
To the best of authors' knowledge, the existing literature in the field of mixed RF/FSO relaying systems have ignored the impact of interference and IQI on the RF links. Effect of hardware impairment has been analyzed on mixed RF/FSO OWR system by the authors in \cite{G} but the impact of mismatch between I and Q components and non-zero boresight component has been neglected. Authors in \cite{mypaper4} have presented the the impact of interference and boresight pointing error for OWR systems, but the impact of IQI has not been addressed. Mixed RF/FSO TWR system affected by co-channel interference (CCI) has been presented in \cite{mypaper3} however, RF front-end has been considered to be ideal. In addition, boresight component has also been excluded from the analysis. \par
Nevertheless, as per the knowledge of authors, no study has been reported on the effect of IQI on RF front-end for interference limited mixed RF/FSO TWR systems. Motivated by this, a unified performance analysis of mixed RF/FSO TWR system is pursued, where the IQI affects both transmitter and receiver of RF front-end. Specifically, following are the key contributions: 1) The statistical model for FSO link undergoing double generalized Gamma (D-GG) turbulence in the presence of non-zero boresight pointing error has been derived. 2) Significantly, the impact of IQI impairments on the performance of mixed RF/FSO TWR system has been quantified. 3) Additionally, the impact of multiple interferers has been included in the analysis. 4) Analytical closed-form expressions have been derived for outage probability and achievable sum rate (ASR) for the proposed model. \par  
The remainder of this paper is organized as follows: Section II describes system and channel models and derives the closed-form probability density function (PDF) of D-GG distribution with non-zero boresight pointing error. In Section III, the expression for outage probability and ASR are derived. Finally, some numerical results are demonstrated in Section IV before concluding the paper in Section V.
\section{System and Channel Models}
\subsection{System Model}
Consider a mixed RF/FSO TWR communication network, where the two source nodes communicate with each other through an AF relay node. In this setup, it is assumed that source node \(S_1\) is a RF mobile device whereas the source node \(S_2\) is an FSO terminal equipped with single telescope. During the bidirectional operation, transmitter and receiver hardwares of the RF node \(S_1\) are assumed to undergo IQI impairments. The IQI coefficients \(K_{\scaleto{\mathrm{1}}{3 pt}}^{\mathrm{t}}\) and \(K_{\scaleto{\mathrm{2}}{3 pt}}^{\mathrm{t}}\) RF front-end are further expressed as \cite{F} \(K_{\scaleto{\mathrm{1}}{3 pt}}^{\mathrm{t/r}} = \frac{1}{2}\left(1+\epsilon^{\mathrm{t/r}}e^{\pm j\phi^{t/r} } \right)\) and \(K_{\scaleto{\mathrm{2}}{3 pt}}^{\mathrm{t/r}} = \frac{1}{2}\left(1-\epsilon^{\mathrm{t/r}}e^{\mp j\phi^{t/r}} \right)\), 
where \(\phi^{t/r} \) and \(\epsilon^{t/r} \) quantify the phase and amplitude mismatch between the transmitter and receiver hardwares, respectively while the positive and negative signs in these relations account for up-conversion and down-conversion, respectively \cite{F}. The superscripts \(t\) and \(r\) denote Tx and Rx, respectively. After optical-to-electrical conversion, the signal received at the relay node when the transmitter of node \(S_1\) operation is impaired by IQI can be given as
$
y_{R, T_1}  =  h_{\scaleto{\mathrm{RF}}{3 pt}} K_{\scaleto{\mathrm{1}}{3 pt}}^{\mathrm{t}} x_1 + h_{\scaleto{\mathrm{RF}}{3 pt}}^{*} K_{\scaleto{\mathrm{2}}{3 pt}}^{\mathrm{t}} x_1^* + {(\eta I)}^{\rho/2} x_2  
+ \sum_{i=1}^{N} h_{\scaleto{\mathrm{i}}{3 pt}} w_{\scaleto{\mathrm{i}}{3 pt}} + n{\scaleto{T_1}{3 pt}}, $
where \(h_{\scaleto{\mathrm{RF}}{3 pt}}\) denotes the channel coefficient on the RF link from source node on the optical link, the average power is defined as \(\mathbb{E}[|x_1|^2] = P_1 \) where \(\mathbb{E}[.]\) denotes the expectation operator and \(\eta \) is the optical-to-electrical conversion ratio, \(n{\scaleto{T_1}{3 pt}}\) denotes the additive white Gaussian noise (AWGN) noise on the source-to-relay link with zero mean and variance \(\sigma^2_{n1} \). The irradiance fluctuation is denoted by \(I\) to transmit information symbol \(x_2\) while the constant \(\rho\) denotes the type of demodulation scheme employed where \(\rho=2\) corresponds to intensity modulation direct detection (IM/DD) demodulation and \(\rho=1\) is applicable to coherent demodulation. Further to this, it is assumed that the relay node operates in the presence of \(N\) CCIs with \(h_{\scaleto{\mathrm{i}}{3 pt}}\) denoting the fading coefficient of the \(i^{th}\) interferer transmitting a symbol \(w_i\) and interference power \(P_{Ii}\) is given by the relation \(\mathbb{E}[|w_i|^2] = P_{Ii} \).  \par
In the second phase of communication, the relay node forwards the information symbol received from node \(S_1\) to the node \(S_2\) and vice-versa. It is assumed that the transmission from both the nodes provide self-interference cancellation. After removing the DC component, the received signal at source node \(S_1\) in the presence of IQI affected receiver is given by
\begin{align*} %(Eq. 4)
y_{\scaleto{\mathrm{S_1, T_2}}{5 pt}} & = G_1h_{\scaleto{\mathrm{RF}}{3 pt}} ({\eta I})^{\rho/2} K_{\scaleto{\mathrm{1}}{3 pt}}^{\mathrm{r}} x_2 
+ G_1h_{\scaleto{\mathrm{RF}}{3 pt}}^{*} ({\eta I})^{\rho/2} K_{\scaleto{\mathrm{2}}{3 pt}}^{\mathrm{r}} x_2^*
+ (K_{\scaleto{\mathrm{1}}{3 pt}}^{\mathrm{r}}+K_{\scaleto{\mathrm{2}}{3 pt}}^{\mathrm{r}})\Big(\sum_{i=1}^{N} h_{\scaleto{\mathrm{i}}{3 pt}} w_{\scaleto{\mathrm{i}}{3 pt}} + N_{01}\Big) .   \addtag \label{sysEq2}
\end{align*}
Similarly, the signal at source node \(S_2\) can be written as
\begin{align*} %(Eq. 5)
y_{\scaleto{\mathrm{S_2, T_2}}{5 pt}} & = G_2 h_{\scaleto{\mathrm{RF}}{3 pt}} ({\eta I})^{\rho/2} K_{\scaleto{\mathrm{1}}{3 pt}}^{\mathrm{t}} x_1
+ G_2 h_{\scaleto{\mathrm{RF}}{3 pt}}^{*} ({\eta I})^{\rho/2} K_{\scaleto{\mathrm{2}}{3 pt}}^{\mathrm{t}} x_1^*
+ \sum_{i=1}^{N} h_{\scaleto{\mathrm{i}}{3 pt}} w_{\scaleto{\mathrm{i}}{3 pt}} + N_{02},   \addtag \label{sysEq4}
\end{align*}
where  \(N_{\scaleto{\mathrm{01}}{3 pt}}\) and \(N_{\scaleto{\mathrm{02}}{3 pt}}\) denote noise contribution in the second time-slot with zero mean and variances \({\sigma^2_{n1}} \) and \({\sigma^2_{n2}} \), respectively. 
From (\ref{sysEq2}), the signal-to-interference-plus-noise-ratio (SINR) can be given as
\begin{align*} %(Eq. 3)
\gamma_{\scaleto{\mathrm{S_1, T_2}}{5 pt}} =  \frac{G_1^2 (\eta I)^\rho |h_{\scaleto{\mathrm{RF}}{3 pt}}|^2     |K_{\scaleto{\mathrm{1}}{3 pt}}^{\mathrm{r}}|^2 E_{\scaleto{\mathrm{S2}}{3 pt}} } { G_1^2 (\eta I)^\rho |h_{\scaleto{\mathrm{RF}}{3 pt}}|^2     |K_{\scaleto{\mathrm{2}}{3 pt}}^{\mathrm{r}}|^2 E_{\scaleto{\mathrm{S2}}{3 pt}} 
	+ G_1^2 |h_{\scaleto{\mathrm{RF}}{3 pt}}|^2 \sum_{i=1}^{N} |h_{i}|^2 E_i + {\sigma^2_{n1}}
}.\addtag \label{sysEq3}
\end{align*}
Since the signal is coming on the RF link, the gains \(G_1\) and \(G_2\) are assumed to be inverse of the respective channel gains in the first time slot and hence the SINR can be written as
\begin{align*} 
\gamma_{\scaleto{\mathrm{S_1, T_2}}{5 pt}}& \approx
\frac{\gamma_{\scaleto{\mathrm{FSO}}{3 pt}}} { \frac{\gamma_{\scaleto{\mathrm{FSO}}{3 pt}}}{\kappa^{\mathrm{r}}} 	+ {\Big(1+ \frac{1}{\kappa^{\mathrm{r}}} \Big)   }{\gamma_{\scaleto{\mathrm{I}}{3 pt}}}},
\addtag \label{j1}
\end{align*}
where \(\kappa^{\mathrm{t/r}} = \frac{|K_{\scaleto{\mathrm{1}}{3 pt}}^{\mathrm{t/r}}|^2}{|K_{\scaleto{\mathrm{2}}{3 pt}}^{\mathrm{t/r}}|^2} \) denotes the  image rejection ratio (IRR) due to hardware impairment \cite{F}. The instantaneous interference-to-noise-ratio (INR) is stated as \( \gamma_{\scaleto{\mathrm{I}}{3 pt}} = \sum_{i=1}^{N} \gamma_{\scaleto{\mathrm{I, i}}{3 pt}} \), where \(\gamma_{\scaleto{\mathrm{I, i}}{3 pt}} = \bar\gamma_{\scaleto{\mathrm{I, i}}{3 pt}} \sum_{i=1}^{N} \frac{|h_{i}|^2}{{\sigma^2_{n2}}} \).   The electrical SNR on the FSO channel can be expressed as \(\gamma_{\scaleto{\mathrm{FSO}}{3 pt}} = {(\eta I)^\rho}/({{\sigma^2_{n2}}}) \).  
Similarly, at the FSO node, the overall SINR can be expressed as
\begin{align*} %(Eq. 8)
\gamma_{\scaleto{\mathrm{S_2, T_2}}{5 pt}}& = 
\frac{1 } { \frac{1}{\kappa^{\mathrm{t}}} 	+ \frac{\gamma_{\scaleto{\mathrm{I}}{3 pt}}}{\gamma_{\scaleto{\mathrm{RF}}{3 pt}} |K_{\scaleto{\mathrm{1}}{3 pt}}^{\mathrm{t}}|^2}  }, \label{j12}
\addtag
\end{align*}
where \( \gamma_{\scaleto{\mathrm{RF}}{3 pt}} = \bar \gamma_{\scaleto{\mathrm{RF}}{3 pt}} |h_{\scaleto{\mathrm{RF}}{3 pt}}|^2  \) and \(\bar \gamma_{\scaleto{\mathrm{RF}}{3 pt}} = \frac{P_1}{\sigma^2_{n2}} \).
\subsection{RF Channel Model}
The RF link has been modeled as \(K\) distributed fading to consider the combined effect of short term fading and shadowing, the PDF of which can be expressed as \cite{K-dist}
\[f_{\gamma_{\scaleto{\mathrm{RF}}{3 pt}}} (\gamma) = A_1  \gamma^{\frac{v}{2}} K_{v}\Bigg({\frac{\sqrt{\gamma}}{\sqrt{\bar \gamma_{\scaleto{\mathrm{RF}}{3 pt}}}a}} \Bigg),  \addtag \label{RF} \]
where \(a\) and \(v\) are the model parameters, \( A_1 = \frac{1}{(2a)^{v+2} \Gamma(v+1) {\bar \gamma_{\scaleto{\mathrm{RF}}{3 pt}}}^{\frac{v+2}{2}} } \) while \(\bar\gamma_{\scaleto{\mathrm{RF}}{3 pt}}\) is the average SNR on the RF link. 
\begin{align*}
F_{ \gamma_{\scaleto{\mathrm{RF}}{3 pt}}} (\gamma)= 1- A_1 \gamma^{\frac{v+2}{2}} 
\text{G}_{0,2}^{2, 0} \Bigg[ \frac{\gamma}{A_2} \Bigg|
\begin{array}{cc}
\noindent\rule{0.4cm}{0.5pt} \\
\frac{v}{2}, \frac{-v}{2}
\end{array}
\Bigg],  \addtag \label{RF-CDF}
\end{align*}
where \(A_2 = 4 \bar \gamma_{\scaleto{\mathrm{RF}}{3 pt}} a^2 \) and \(G[.] \) represents the Meijer-G function defined in \cite[Eq. (9.301)]{RyzhikTables}. Assuming that the fading caused by the interference, \( h_{\scaleto{\mathrm{I}}{3 pt}}\), is Nakagami-\(m\) distributed with \( \bar\Omega_{\scaleto{\mathrm{I}}{3 pt}} = \frac{\mathbb{E} [|h_{\scaleto{\mathrm{i}}{3 pt}}|^2]}{\sigma^2_n} \), the corresponding PDF of \( \gamma_{\scaleto{\mathrm{I}}{3 pt}}\) can be written as \cite{Nak-Intf}:
\begin{align*}
f_{ \gamma_{\scaleto{\mathrm{I}}{3 pt}}} (x) = \left(\frac{m_{\scaleto{\mathrm{I}}{3 pt}}}{ \bar\Omega_{\scaleto{\mathrm{I}}{3 pt}}}\right)^{N m_{\scaleto{\mathrm{I}}{3 pt}}} \frac{x^{Nm_{\scaleto{\mathrm{I}}{3 pt}}-1}}{\Gamma(N m_{\scaleto{\mathrm{I}}{3 pt}})} e^{-\frac{m_{\scaleto{\mathrm{I}}{3 pt}}}{ \bar\Omega_{\scaleto{\mathrm{I}}{3 pt}}} x}. \addtag \label{intf-PDF}
\end{align*}
\subsection{FSO Channel Model}
The irradiance fluctuation on the FSO link is modeled as \(I = I_a I_p \), where \(I_a\) denotes the atmospheric turbulence with \(I_a=I_xI_y\), such that \(I_x\sim GG(\alpha_1, \beta_1, \Omega_1) \), and \(I_y\sim GG(\alpha_2, \beta_2, \Omega_2) \) \cite{DGGwithPointError}, whereas \(I_p\) accounts for the pointing error. The PDF of \(I_a\) on the FSO link can be expressed as \cite{DGGwithPointError}:
\begin{align*}%(Eq. 16)
\setcounter{equation}{8}
f_{{\scaleto{\mathrm{I_a}}{5 pt}}}(I_a) = \frac{{D}_1}{I_a} \meijerG{0}{\lambda+\sigma}{\lambda+\sigma}{0}{1-\tau_0}{\noindent\rule{0.4cm}{0.5pt}}{\frac{{D}_2}{I_a^y}  }, \addtag \label{FSO-PDF}
\end{align*}
where \(\mathcal{D}_1 = \frac{y \sigma^{\beta_1-\frac{1}{2}} \lambda^{\beta_2-\frac{1}{2}} (2\pi)^{1-\frac{\sigma+\lambda}{2}} }{\Gamma(\beta_1) \Gamma(\beta_2)} \), 
\( \mathcal{D}_2 = \frac{\lambda^\lambda \sigma^\sigma \Omega_1^{\sigma} \Omega_2^{\lambda}}{\beta_1^\sigma \beta_2^\lambda} \), \(\tau_0 = \left[ \Delta(\sigma:\beta_1), \Delta(\lambda:\beta_2) \right] \) \cite{DGGwithPointError} with \(\Delta(z:x)\) defined as \([\frac{x}{z}, \frac{x+1}{z}, \dots, \frac{x+z-1}{z}]\). Moreover, constant \(y=\alpha_2\lambda \). The PDF of pointing error with non-zero boresight component can be defined as \cite{A}
\begin{align*}
f_{I_p}(I_p) & = \frac{\xi^2}{A_0^{\xi^2}} \exp\left[-\frac{b^2}{2 \sigma_s^2}\right] I_p^{\xi^2-1} 
I_{0}\Bigg(\frac{b}{\sigma_s^2} \sqrt{\frac{-w_{z_{eq}}^2 \text{ln}\frac{I_p}{A_0}}{2} } \Bigg), 0<I_p<A_0, \addtag \label{j3}
\end{align*}
%%%%%%%%%%%%%%%%%%%%%%%%%%%%% To be parapharsed %%%%%%%%%%%%%%%%%%%%%%%%%%%%%%%%%%%%%
where \(b\) denotes the boresight displacement parameter and \(I_0(.) \) denotes the modified Bessel function of the first kind and zero order defined in \cite[eq. (8.431.1)]{RyzhikTables}. For \(b \to 0 \), the pointing error model in (\ref{j3}) specializes for the case of zero-boresight error. The various pointing error parameters involved are specified as: \(\xi=\frac {w_{z_{eq}}}{2\sigma_s} \) \cite{A}, \(w_{z_{eq}}^2 = \frac {w_{z}^2 \sqrt{\pi} erf(v)}{2 v \exp (-v^2)}\), \(v = \frac{\sqrt \pi r}{2 w_z}\), and \(A_0 = [\text{erf(v)}]^2\) where \(w_z\) is the beam waist (calculated at \(e^{-2}\)) of the Gaussian spatial beam profile and \(w_{z_{eq}}\) is equivalent beam waist at a distance of \(z\). 
%%%%%%%%%%%%%%%%%%%%%%%%%%%%% To be parapharsed %%%%%%%%%%%%%%%%%%%%%%%%%%%%%%%%%%%%%
\\
{\textit{Theorem 1:}} The PDF on the FSO link undergoing D-GG turbulence with non-zero boresight pointing error can be derived as
\begin{align*}
& f(I)  =  \frac{D_1 \xi^2 e^{\big(-\frac{b^2}{2 \sigma_s^2}\big)}}{yI}  \sum_{m=0}^{n} B_{m, n}  
\frac{\partial^{m}}{\partial s^{m}} 
\text{G}_{t_0, y}^{0, t_0}\left[ {\frac{D_2 A_0^y}{I^{y}}}
\middle\vert
\begin{array}{c}\noalign{\vskip-5pt}
\tau_{2} \\ \noalign{\vskip-5pt}
\tau_{1} 
\end{array}
\right]. \addtag \label{j15}
\end{align*}
The constant \(B_{m, n} = \hat{b}_{m, n}  \Big(\frac{\sqrt{2} b w_{z_{eq}}}{\sigma_s}\Big)^{2m} \) with\\ \(\hat{b}_{m, n} = (-1)^{m+1} \frac{(n+m-1)! n^{1-2m}}{m! (n-m)! \Gamma(m+1)} \), \(\tau_1=\Delta(y, s)\), \(\tau_2=\Delta(y,s+1), \Delta(1, 1-\tau_0)\), \(t_0 = \lambda+\sigma+y \) and \(s=-\xi^2 \). \\
\textit{Proof:} See Appendix A.     
\\
{\textit{Lemma 1:}} The cumulative distribution function (CDF) on the FSO channel over D-GG atmospheric turbulence with non-zero boresight pointing error can be derived as follows
\begin{align*}%(Eq. 10)
& F_{\gamma_{{\scaleto{\mathrm{FSO}}{2 pt}}}}(\gamma)={D_3} \gamma^{\frac{r-1}{2}} \sum_{m=0}^{n} B_{m, n}  \frac{\partial^{m}}{\partial s^{m}} \meijerG{t_1}{y}{y(r+1)}{t_1+y}{\tau_4}{\tau_3}{{D_4} {\gamma}^{y} }, \addtag \label{FSO-CDF}
\end{align*}
% t1=k3, t2=k2, t1'=1-k3=k5, t2'=1-k2=k4, t3=k6, t4=k7
where \(t_1=\rho t_0\), \( D_3 = \frac{D_1 {\xi^2} e^{({-\frac{b^2}{2 \sigma_s^2}})}  \rho^{\delta-1}}{y^2( 2\pi)^{(0.5(\rho-1)(\lambda+\sigma-y))}} \), \(\bar \tau_{i}=1-\tau_i \) with \(\delta = \sum_{i=1}^{\lambda+\sigma+y}\bar \tau_{2, i}- \sum_{i=1}^{y}\bar \tau_{1, i}+\frac{2-\lambda-\sigma}{2}\),\\ \(D_4 = \frac{D_2 ({A_0 \mathbb{E}[I]})^{y}}{{{\mu_\rho}^{y/\rho}}} \) and \(\mu_\rho = \frac{(\eta I)^\rho}{\sigma_{n}^2} \). The parameters involved are formulated as: \(\tau_4 = \left[\Delta(\rho:\bar \tau_1), \Delta(y:\frac{1-\rho}{\rho})\right]\) and \(\tau_3= \left[ \Delta(y:1-\frac{\rho-1}{\rho}), \Delta(\rho:\bar \tau_2) \right] \). \\
\textit{Proof:} With the aid of \cite[Eq. (7.8.11.4)]{RyzhikTables}, the closed-form expression for the \(\mathbb{E}[I]\) can be obtained from (\ref{j15}). Further to this, after performing random variable transformation using \\
\(I = \frac{\gamma^{1/\rho}}{\mu_\rho} \) and applying \cite[Eq. (07.34.21.0084.01)]{Wolfram}, the required expression for the CDF can be derived.
\section{Performance Analysis}
\subsection{Exact Outage Probability Analysis}
The outage probability quantifies the reliability of wireless communication system and is defined as the probability that the instantaneous SNR of the link falls below a predefined threshold value \({\gamma_{\scaleto{\mathrm{th}}{3 pt}}} \).\\
{\textit{Theorem 2:}} The outage probability of the interference-limited mixed RF/FSO TWR with hardware impairment can be expressed in closed-form in (\ref{j23}). \\
The parameter \(\tau_5= \Delta(y:(1-\frac{\rho-1}{\rho}-m_{\scaleto{\mathrm{I}}{3 pt}}N)) \) .\\
%%%%%%%%%%%%%%%%%%%%%%%%%%%%%%%%%%%%%%%%%%%%%%%%%%%%%%%%%%%%%%%%%%%%%%%%%%%%%%%%%%%%%%%%
%%%%%%%%%%%%%%%%%%%%%%%%%%%%%%%%%%%%%%%%%%%%%%%%%%%%%%%%%%%%%%%%%%%%%%%%%%%%%%%%%%%%%%%%
\begin{figure*}[!t] % Eq (17) Outage Probability
	\setcounter{equation}{12}
\small	\begin{align*}
	& P_{\mathrm{out}} = 1- \Bigg[\Bigg(
	\Bigg(\frac{\bar \Omega_{\scaleto{\mathrm{I}}{3 pt}}}{m_I} \Bigg)^{\frac{v+2}{2}}
	\frac{A_1}{\Gamma(m_IN)}
	\Bigg[ \frac{\gamma_{\scaleto{\mathrm{th}}{3 pt}}}{ |K_{\scaleto{\mathrm{1}}{3 pt}}^{\mathrm{t}}|^2 (1-\frac{\gamma_{\scaleto{\mathrm{th}}{3 pt}}}{\kappa^{\mathrm{t}}}) } \Bigg]^{\frac{v+2}{2}} 
	\text{G}_{1,2}^{2, 1} \Bigg[ \frac{\gamma_{\scaleto{\mathrm{th}}{3 pt}}{\bar \Omega_{\scaleto{\mathrm{I}}{3 pt}}}}{ A_2 m_I |K_{\scaleto{\mathrm{1}}{3 pt}}^{\mathrm{t}}|^2 { (1-\frac{\gamma_{\scaleto{\mathrm{th}}{3 pt}}}{\kappa^{\mathrm{t}}}) } }
	\Bigg|
	\begin{array}{cc}
	1-(m_IN)-\frac{v+2}{2} \\
	-\frac{v+2}{2}, \frac{v}{2}
	\end{array}
	\Bigg]
	\Bigg)
	\\ & \times %%%%%%%%%% FSO Link
	\Bigg(1- \frac{D_3 y^{(\frac{\rho-1}{\rho}+m_{\scaleto{\mathrm{I}}{3 pt}}N-\frac{1}{2})}}{(2\pi)^{\frac{y-1}{2}}\Gamma(m_{\scaleto{\mathrm{I}}{3 pt}}N)}  
	\Big\{ \frac{\bar\Omega_{\scaleto{\mathrm{I}}{3 pt}} (\kappa^{\mathrm{r}}+1)\gamma_{\scaleto{\mathrm{th}}{3 pt}}}{m_{\scaleto{\mathrm{I}}{3 pt}}(\kappa^{\mathrm{r}}-\gamma_{\scaleto{\mathrm{th}}{3 pt}})} \Big\}^{\frac{\rho-1}{\rho}} \sum_{m=0}^{n} B_{m, n} 
	\frac{\partial^{m}}{\partial s^{m}}
	\meijerG{t_1}{2y}{2y+\rho y}{t_1+y}{\tau_5}{\tau_3}{D_4 \left(\frac{y \bar\Omega_{\scaleto{\mathrm{I}}{3 pt}}}{m_{\scaleto{\mathrm{I}}{3 pt}} }\right)^{y}   \Big\{ \frac{(\kappa^{\mathrm{r}}+1)\gamma_{\scaleto{\mathrm{th}}{3 pt}}}{(\kappa^{\mathrm{r}}-\gamma_{\scaleto{\mathrm{th}}{3 pt}})} \Big\}^y } \Bigg)\Bigg].
	\addtag \label{j23}
	\end{align*}
	\rule{\linewidth}{1pt}
\end{figure*}
%%%%%%%%%%%%%%%%%%%%%%%%%%%%%%%%%%%%%%%%%%%%%%%%%%%%%%%%%%%%%%%%%%%%%%%%%%%%%%%%%%%%%%
%%%%%%%%%%%%%%%%%%%%%%%%%%%%%%%%%%%%%%%%%%%%%%%%%%%%%%%%%%%%%%%%%%%%%%%%%%%%%%%%%%%%%%
\textit{Proof:} See Appendix B.     \\
The closed-form expression in (\ref{j23}) shows that the outage probability depends on the IQI due to hardware imperfections of the RF source. In addition, it is affected by the shadowing and fading effects of RF link and strength and number of interferers at the relay node. Moreover, atmospheric turbulence of FSO link and non-zero boresight pointing error also impacts the behavior of the outage probability.

\subsection{Asymptotic Outage Probability Analysis}
We now present an asymptotic expression for the exact outage probability in (14). Consider a high SNR regime where (\(\bar \gamma_{\scaleto{\mathrm{RF}}{3 pt}},\mu_\rho \to \infty\)), the outage probability can be approximated as \(P_{\mathrm{out}} \simeq  F_{\gamma_{\scaleto{\mathrm{S_1, T_2}}{5 pt}}}(\gamma_{\scaleto{\mathrm{th}}{3 pt}})+F_{\gamma_{\scaleto{\mathrm{S_2, T_2}}{5 pt}}}(\gamma_{\scaleto{\mathrm{th}}{3 pt}})\). Moreover, as \(\bar \gamma_{\scaleto{\mathrm{RF}}{3 pt}} \to \infty \), the value of \(A_2 \to 0 \) and similarly, when \(\mu_\rho \to \infty \), the magnitude of \(D_4\) becomes arbitrarily small. By virtue of \cite[Eq. (07.34.06.0006.01)]{Wolfram}, for small argument values, the Meijer-G functions can be expressed in terms of elementary functions with the aid of dominant poles \cite[Theorem (1.1), (1.2)]{ResTheom}, thus providing the asymptotically approximated expressions as
\begin{align*}
\setcounter{equation}{13}
& P_{\mathrm{out}} \simeq 1-  \frac{A_1 \Theta_1(v)}{\Gamma(m_IN)}
\Bigg[ \frac{\gamma_{\scaleto{\mathrm{th}}{3 pt}} \bar \Omega_{\scaleto{\mathrm{I}}{3 pt}}}{ |K_{\scaleto{\mathrm{1}}{3 pt}}^{\mathrm{t}}|^2 (1-\frac{\gamma_{\scaleto{\mathrm{th}}{3 pt}}}{\kappa^{\mathrm{t}}}) m_I} \Bigg]^{\frac{v+2}{2}} + \frac{D_3}{\Gamma(m_{\scaleto{\mathrm{I}}{3 pt}}N)}
\frac{ y^{(\frac{\rho-1}{\rho}+m_{\scaleto{\mathrm{I}}{3 pt}}N-\frac{1}{2})}}{(2\pi)^{\frac{y-1}{2}}}  
\Big\{ \frac{\bar\Omega_{\scaleto{\mathrm{I}}{3 pt}} (\kappa^{\mathrm{r}}+1)\gamma_{\scaleto{\mathrm{th}}{3 pt}}}{m_{\scaleto{\mathrm{I}}{3 pt}}(\kappa^{\mathrm{r}}-\gamma_{\scaleto{\mathrm{th}}{3 pt}})} \Big\}^{\frac{\rho-1}{\rho}} 
\\
& \times
\sum_{m=0}^{n} B_{m, n} 
\frac{\partial^{m}}{\partial s^{m}} \Theta_2(\mathbb{P}), \addtag \label{out-asym}
\end{align*}
where \(\Theta_3=\frac{\gamma_{\scaleto{\mathrm{th}}{3 pt}}{\bar \Omega_{\scaleto{\mathrm{I}}{3 pt}}}}{ A_2 m_I |K_{\scaleto{\mathrm{1}}{3 pt}}^{\mathrm{t}}|^2 { (1-\frac{\gamma_{\scaleto{\mathrm{th}}{3 pt}}}{\kappa^{\mathrm{t}}}) } } \), \(\Theta_1(v) = (\Theta_3^{\frac{-v-2}{2}})
\times \Gamma(v+1) \Gamma(m_I N)+(\Theta_3^{\frac{v}{2}}) \Gamma(-v-1)\\
\times \Gamma(m_IN+v+1)  \), \(\mathbb{P} = \min[\Delta(y:1/\rho)] \) and \(\Theta_2(\mathbb{P}) = \Theta_4^{\mathbb{P}} \frac{\prod_{j=1,\mathbb{P} \neq \tau_{3, j}}^{t_1} \Gamma(\tau_{3, j}- \mathbb{P}) \prod_{j=1}^{2y} \Gamma(1-\tau_{5, j}+\mathbb{P})}{\prod_{j=2y+1}^{2y+\rho y} \Gamma(\tau_{5, j}-\mathbb{P}) \prod_{j=t_1+1}^{t_1+y} \Gamma(1-\Delta(y:\bar \tau_{2, j}+ \mathbb{P}))} \) with \(\Theta_4 = D_4 \left(\frac{y \bar\Omega_{\scaleto{\mathrm{I}}{3 pt}}}{m_{\scaleto{\mathrm{I}}{3 pt}} }\right)^{y}   \Big\{ \frac{(\kappa^{\mathrm{r}}+1)\gamma_{\scaleto{\mathrm{th}}{3 pt}}}{(\kappa^{\mathrm{r}}-\gamma_{\scaleto{\mathrm{th}}{3 pt}})} \Big\}^y \). It is to be noted that the approximate expression reduces the complexity of evaluation.
\subsection{Exact Achievable Sum Rate Analysis}
To further gain deep insight into the performance of system, the ASR offered by the mixed RF/FSO relaying system can be analyzed. Analytically, ASR, \(\mathcal{R} = \mathcal{R}_1 + \mathcal{R}_2 \), where \(\mathcal{R}_i = (1/2) \mathbb{E}[\text{log}_2 (1+ \gamma_{\scaleto{\mathrm{S_i, T_2}}{5 pt}} ) ]  \). \\
{\textit{Theorem 3:}} The exact expression for the ASR offered by the mixed RF/FSO TWR system with I/Q imbalance in the presence of \(N\) CCIs at the relay node can be expressed as
\begin{align*}
\setcounter{equation}{14}
& \mathcal{R} \simeq \frac{1}{2 \text{log}(2)} \Bigg[ D_5 \sum_{m=0}^{n} B_{m, n}
\text{G}_{2y+\rho y, t_1+y}^{t_1+y, y} \Bigg[ \frac{D_6}{\Delta^y}
\Bigg|
\begin{array}{cc}
\tau_6 \\
\tau_7
\end{array}
\Bigg]
 +
\frac{A_1}{\Gamma(m_I N)}\Big(\frac{1}{|K_{\scaleto{\mathrm{1}}{3 pt}}^{\mathrm{t}}|^2} \Big)^{{\frac{v}{2}}+1}
\Bigg(\frac{\bar \Omega_{\scaleto{\mathrm{I}}{3 pt}}}{m_I} \Bigg)^{\frac{v}{2}}
\\ & \times
\text{G}_{3, 2}^{3, 4} \Bigg[\frac{\bar \Omega_{\scaleto{\mathrm{I}}{3 pt}}}{A_2 |K_{\scaleto{\mathrm{1}}{3 pt}}^{\mathrm{t}}|^2 m_I}
\Bigg|
\begin{array}{cc}
\tau_8 \\
\tau_9
\end{array}
\Bigg]
,  \addtag \label{ASR}
\end{align*}
where \(D_5 = \Big(\frac{1+\kappa^{\mathrm{r}}}{\kappa^{\mathrm{r}}} \Big)^{\frac{\rho-1}{\rho}} \frac{D_1 {\xi^2} e^{({-\frac{b^2}{2 \sigma_s^2}})}}{\Gamma(m_I N)}\Bigg(\frac{\bar \Omega_{\scaleto{\mathrm{I}}{3 pt}}}{m_I} \Bigg)^{\frac{1}{\rho}} \Lambda^{\frac{1-\rho}{\rho}}  \times\frac{y^{{m_IN-\frac{1}{\rho}+1}}}{(2\pi)^{\frac{3}{2}(y-1)+\frac{(\rho-1)}{2}(\lambda+\sigma)}} \) and \(D_6=\frac{D_4^\rho}{\rho^{(\rho(\lambda+\sigma))}} \big(\frac{(\kappa^{\mathrm{r}}+1) y\bar \gamma_{\scaleto{\mathrm{I}}{3 pt}}}{m_I \kappa^{\mathrm{r}} \Delta} \big)^y \) with \(\Lambda=1 \quad \text{and} \quad e/{2\pi}\) when \(r=1\) and \(r=2\), respectively. The derived parameters can be defined as \\  \( \tau_{6} = [\Delta(y:(1-m_IN+\frac{1}{\rho})), \Delta(y:(1-\frac{\rho-1}{\rho})), \Delta(\rho:\bar \tau_1)] \), \(\tau_{7} = [\Delta(\rho:\bar \tau_2), \Delta(y:(\frac{1-\rho}{\rho}))\), \(\tau_{8} = [1-(m_IN+\frac{v}{2}), -\frac{(v+2)}{2}, 1-\frac{(v+2)}{2}] \), and \(\tau_9=[\frac{v}{2}, -\frac{v}{2}, -\frac{(v+2)}{2}, -\frac{(v+2)}{2}] \).  \\
\textit{Proof:} See Appendix C.      \\
The exact ASR expression accounts for source Tx/Rx IQI, CCIs at relay node, fading and shadowing on the RF link, and non-zero boresight pointing error along with atmospheric turbulence on the FSO link. Applying the same steps described in Section B, the asymptotic expression of ASR can be obtained as
\begin{align*}
\mathcal{R} \approx & \frac{1}{2 \text{log}(2)}  D_5 \sum_{m=0}^{n} B_{m, n}
\Theta_5(\mathbb{Q})+ \frac{A_1}{{|K_{\scaleto{\mathrm{1}}{3 pt}}^{\mathrm{t}}|^{v+2}}}
\Bigg(\frac{\bar \Omega_{\scaleto{\mathrm{I}}{3 pt}}}{m_I} \Bigg)^{\frac{v}{2}}
\Gamma(v) \left(\frac{\bar \Omega_{\scaleto{\mathrm{I}}{3 pt}}}{A_2 |K_{\scaleto{\mathrm{1}}{3 pt}}^{\mathrm{t}}|^2 m_I}\right)^{\frac{-v}{2}}
\addtag \label{j40}
\end{align*}
where \(\Theta_5(\mathbb{Q}) =  \frac{\Theta_4^{\mathbb{Q}} \prod_{j=1,\mathbb{Q} \neq \tau_{7, j}}^{t_1+y} \Gamma(\tau_{7, j}- \mathbb{Q}) \prod_{j=1}^{y} \Gamma(1-\tau_{6, j}+\mathbb{Q})}{\prod_{j=y+1}^{2y+\rho y} \Gamma(\tau_{6, j}-\mathbb{Q}) } \) with \(\mathbb{Q} =\min(\tau_{7}) \).
\section{Results}
\begin{figure}[!ht]
	\centering
	\includegraphics[width=8cm, height=6cm]{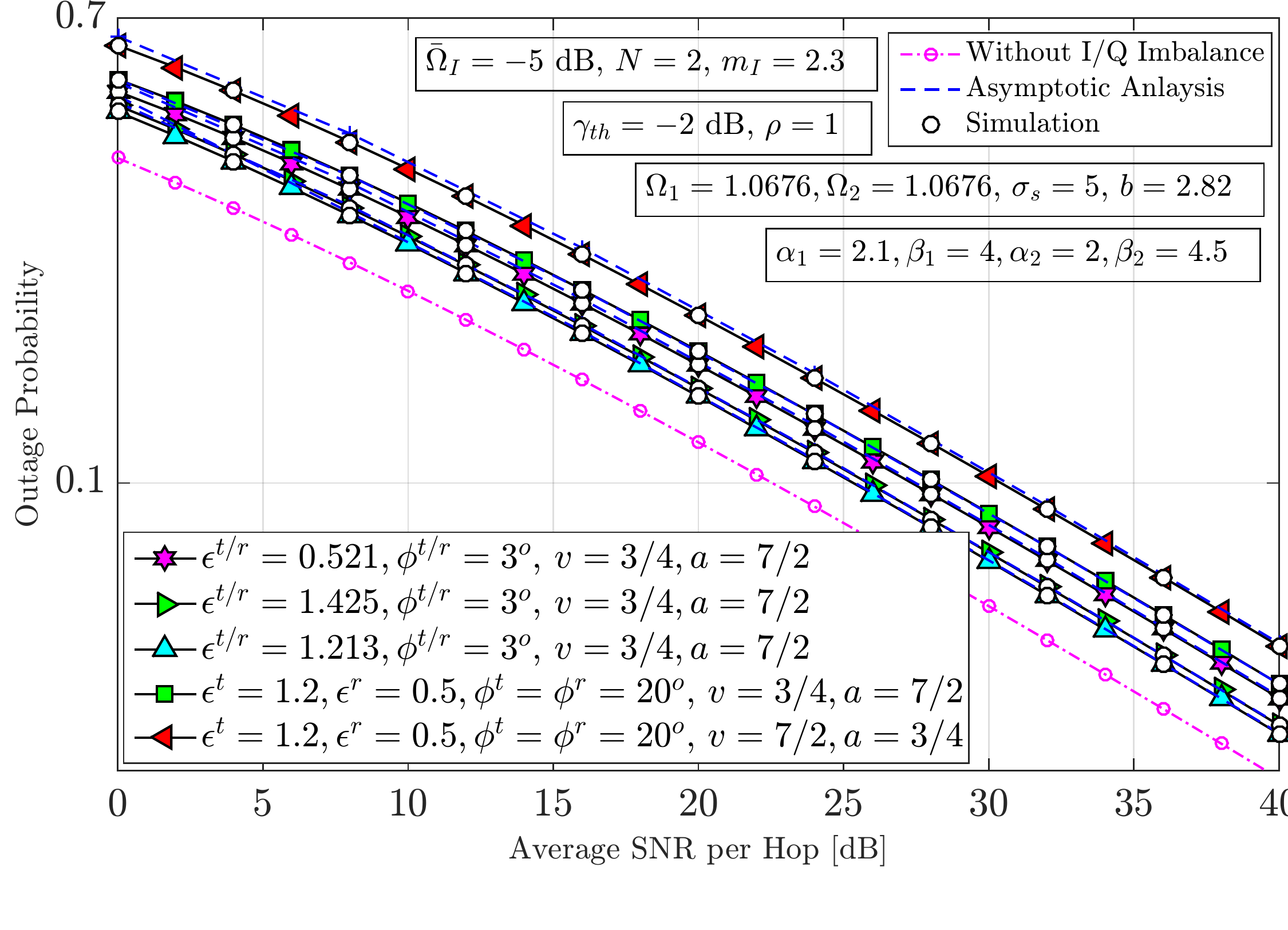}
	\caption{\small{Effect of RF Tx/Rx IQI on outage performance of mixed RF/FSO TWR system.}}\label{out11}
\end{figure}

In this section, numerical results have been presented to quantify the findings by considering same average SNRs of both the hops. The effect of IQI at RF front-end on the outage performance has been demonstrated in Fig. \ref{out11}. The outage probability comparison is shown for different IRR values of \( \kappa^{t/r} = \{10, 15, 20\}\) dB with phase error \( {\phi^t = 3^{\circ} } \) and \(\epsilon^{\mathrm{t/r}}=\{0.521, 1.425, 1.213\}\). It can be noted from the plot that as the value of IRR, \(\kappa^{t/r}\) increases, the reliability of considered system improves. In addition, plots in Fig. 1 demonstrate the impact of IQI imbalances with large phase error \( {\phi^t = 20^{\circ} } \), which is further compared with the results of without IQI impairment. The impact of IQI impairment is imperative in the characteristic curve. Moreover, the fading and shadowing on the RF link are considered to be \(a=\{3/4, 7/2\}\) and \(v=\{3/4, 7/2\}\), respectively. It can be inferred from the plot that the exact expressions are validated by their asymptotic expressions for the practical range of SNRs. \par 
%%%%%%%%%%%%%%%%%%%%%%%%%%%%%%%%%% Figure Fixed Gain %%%%%%%%%%%%%%%%%%%%%%%%%%%%%%% %%%%%%%%%%%%%%%%%%%%%%%%%%%%%%%%%%%%%%%%%%%%%%%%%%%%%%%%%%%%%%%%%%%%%%%%%%%%%%%%%%%%
%%%%%%%%%%%%%%%%%%%%%%%%%%%%%%%%%%%%%%%%%%%%%%%%%%%%%%%%%%%%%%%%%%%%%%%%%%%%%%%%%%%%
\begin{figure}[!ht]
	\centering
	\includegraphics[width=8cm, height=6cm]{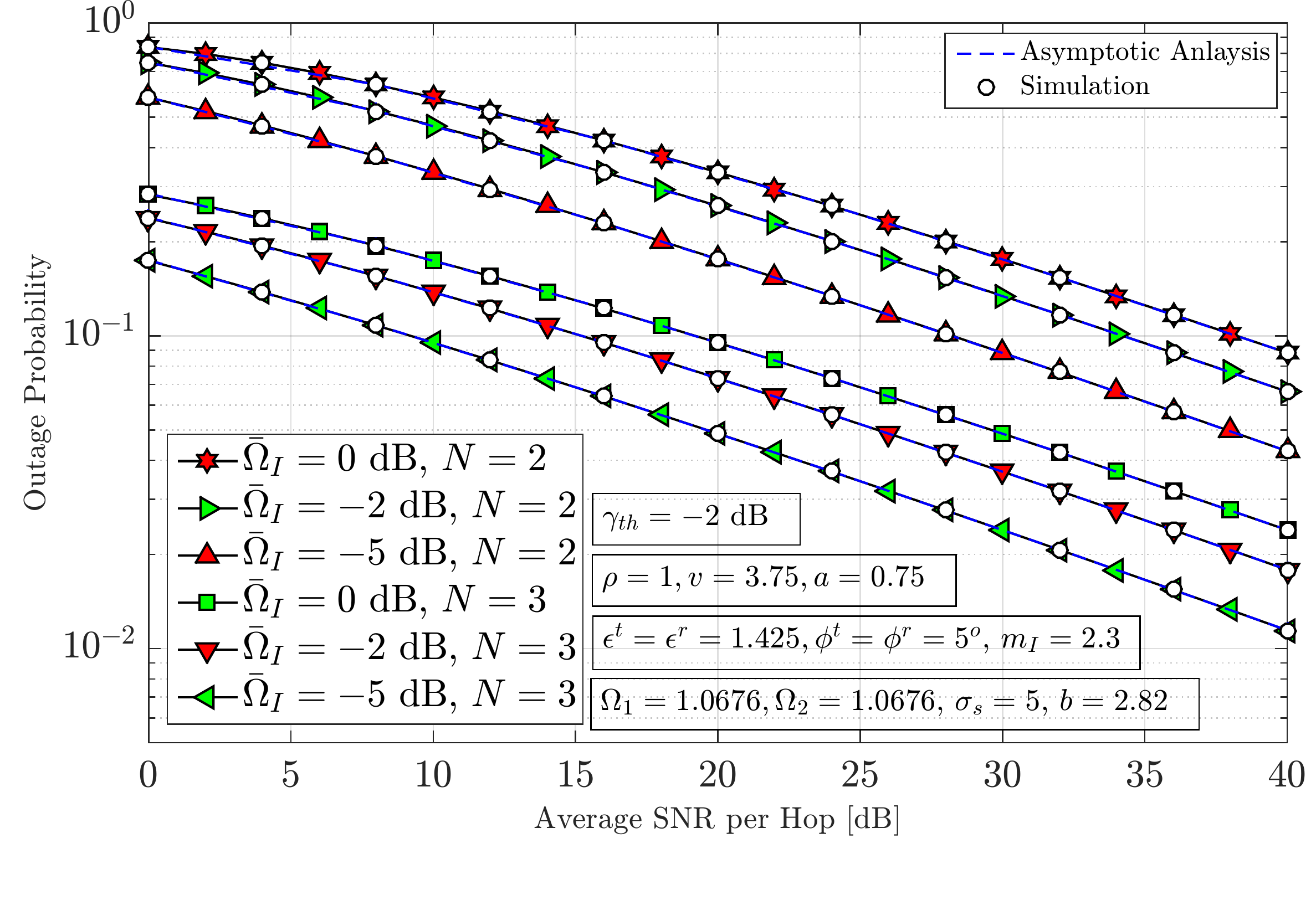}
	\caption{\small{Outage probability against average INR for different interfering signals.}}\label{out22}
\end{figure}
\begin{figure}[!ht]
	\centering
	\includegraphics[width=8cm, height=6cm]{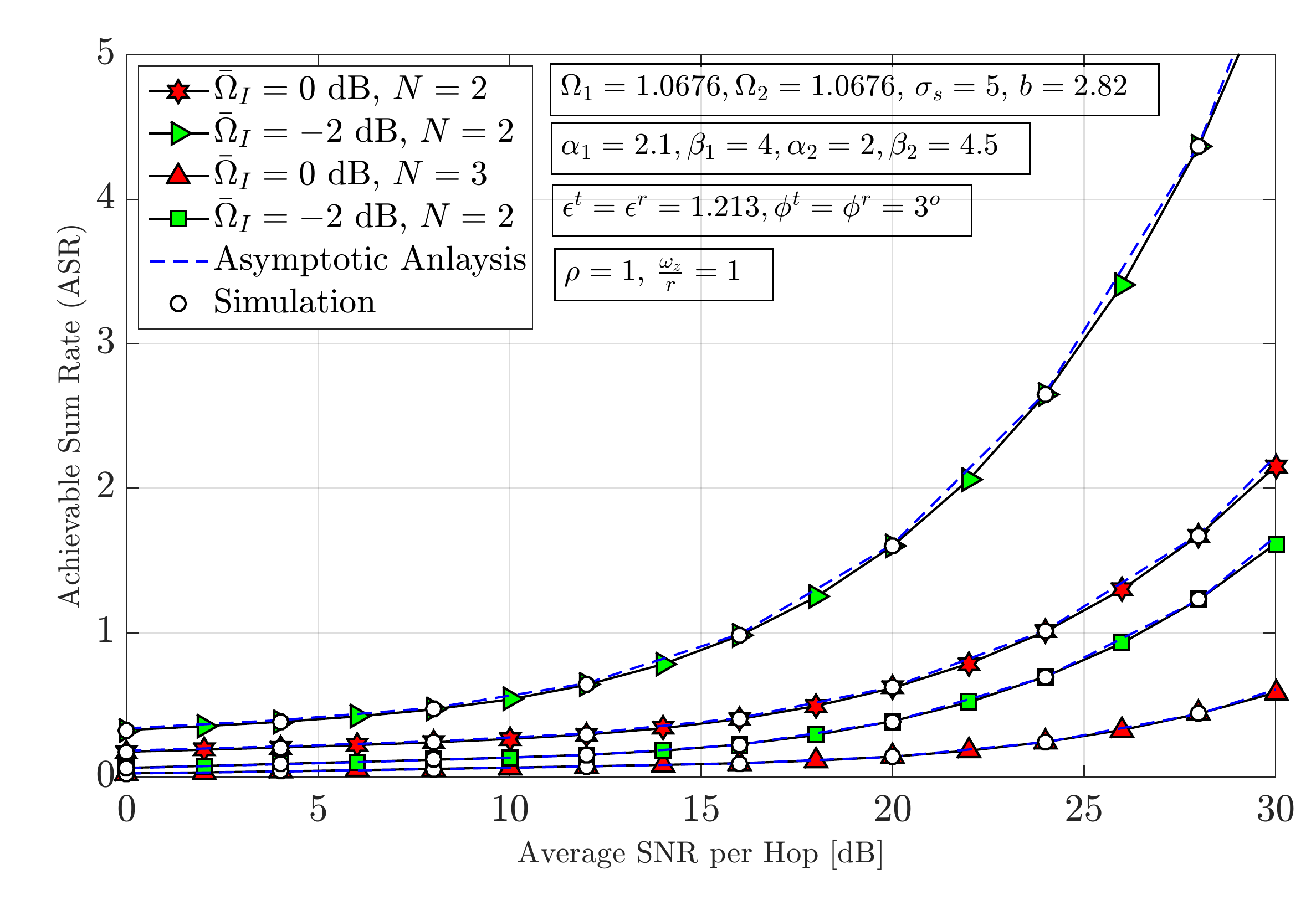}
	\caption{\small{ASR performance by varying number and strength of CCIs at the relay node.}}\label{cap3}
	
\end{figure}
%%%%%%%%%%%%%%%%%%%%%%%%%%%%%%%%%%%%%%%%%%%%%%%%%%%%%%%%%%%%%%%%%%%%%%%%%%%%%%%%%%%%%
%%%%%%%%%%%%%%%%%%%%%%%%%%%%%%%%%%%%%%%%%%%%%%%%%%%%%%%%%%%%%%%%%%%%%%%%%%%%%%%%%%%%%
%%%%%%%%%%%%%%%%%%%%%%%%%%%%%%%%%%%%%%%%%%%%%%%%%%%%%%%%%%%%%%%%%%%%%%%%%%%%%%%%%%%%%
%%%%%%%%%%%%%%%%%%%%%%%%%%%%%%%%%%%%%%%%%%%%%%%%%%%%%%%%%%%%%%%%%%%%%%%%%%%%%%%%%%%%%
%%%%%%%%%%%%%%%%%%%%%%%%%%%%%%%%%%%%%%%%%%%%%%%%%%%%%%%%%%%%%%%%%%%%%%%%%%%%%%%%%%%%%	
In Fig. \ref{out22}, impact of number of interferers and their strengths has been addressed. The average INR considered is \(\bar \Omega_{\scaleto{\mathrm{I}}{3 pt}} = \{0, -2, -5\} \) dB for \(N=\{2, 3\}\) interferers with \(m_I = 2.3\). The reliable transmission can be clearly demonstrated by the rapidly falling curves of outage probability when the strength of the interferers/number of interferes reduces. In Fig. \ref{cap3}, the derived ASR expression has been verified with Monte-Carlo simulations and its asymptotic analysis, where it can be observed that with the increase in the number and/or strength of interference, the ASR of the considered mixed RF/FSO relaying system in the presence of RF IQI and non-zero boresight pointing error reduces. 
\section{Conclusions}
In this letter, we analyzed the outage performance and ASR of RF/FSO TWR based network in the presence of RF Tx/Rx IQI and CCIs. A unified expression for PDF on the FSO link statistics undergoing D-GG atmospheric turbulence with non-zero boresight pointing error has been derived that further serves as an analytical tool to incorporate practicability in the study. In the light of derived exact and asymptotic results, it has been concluded that the imbalance between I and Q components in the RF circuitry along with type of optical demodulation scheme used on the FSO link impact the performance of mixed RF/FSO TWR system. In addition to this, the strength and number of interferers limit the achievable quality of service for the considered system. 
%%%%%%%%%%%%%%%%%%%%%%%%%%%%%%%%%%%%%%%%%%  Appendix %%%%%%%%%%%%%%%%%%%%%%%%%%%%%%%%
%%%%%%%%%%%%%%%%%%%%%%%%%%%%%%%%%%%%%%%%%%%%%%%%%%%%%%%%%%%%%%%%%%%%%%%%%%%%%%%%%%%%
%%%%%%%%%%%%%%%%%%%%%%%%%%%%%%%%%%%%%%%%%%%%%%%%%%%%%%%%%%%%%%%%%%%%%%%%%%%%%%%%%%%%
\appendices
\section{FSO Link with Nonzero Boresight Pointing Error}
For deriving the statistics on the FSO link undergoing D-GG atmospheric turbulence with non-zero boresight pointing error, the integral \(f(I) = \int f_{I/I_a}\Big({I}/{I_a}\Big) f(I_a) dI_a\) is determined. Therefore, substituting the PDFs of turbulence and pointing error into the integral, the resulting expression can be stated as
\begin{align*}
f(I) &= \frac{\xi^2D_1}{ A_o^{\xi^2} } \exp\left(-\frac{b^2}{2\sigma_s^2} \right) I^{\xi^2-1}  \int_{I/A_o}^{\infty}  I_a^{\xi^2-1}  
\meijerG{0}{\lambda+\sigma}{\lambda+\sigma}{1}{1-\tau_0}{\tau_1}{\frac{{D}_2}{I_a^y}  } 
I_0\left(\frac{b}{\sigma^2_s} \sqrt{-2\xi^2 \text{ln}\Bigg(\frac{I}{I_a A_o}\Bigg)} \right) dI_a \addtag \label{fI-boresig}
\end{align*}
Moreover, with the aid of \cite[Eq. (19)]{BesselsApprx}, the Bessel's function \(I_{0}(x)\) can be expressed in summation form, which is substituted in (\ref{fI-boresig}). Thereafter, integral is formulated as
\begin{align*}
f(I) &= \frac{\xi^2D_1}{ A_o^{\xi^2} } \exp\left(-\frac{b^2}{2\sigma_s^2} \right) I^{\xi^2-1}  \sum_{m=0}^{n}  \hat{b}_{m,n,0} \left(\frac{b \xi}{\sqrt{2}\sigma_s}\right)^{2m} \\ & \times \underbrace{\int_{I/A_o}^{\infty}  I_a^{\xi^2-1} 
	\meijerG{0}{\lambda+\sigma}{\lambda+\sigma}{1}{1-\tau_0}{\tau_1}{\frac{{D}_2}{I_a^y}  } 
	{\Bigg\{ \text{ln}\Bigg(\frac{I}{I_a A_o}\Bigg)\Bigg\}}^{m} dI_a}_{J} \addtag 
\end{align*}
After performing some mathematical manipulations, the inner integral \(J\) can be expressed as  $
J = \int_{0}^{\infty} -\Big(\frac{I}{A_o}\Big)^{-\xi^2+1} \\\times e^{t(\xi^2-1)} t^{m} 
\meijerG{0}{\lambda+\sigma}{\lambda+\sigma}{1}{1-\tau_0}{\tau_1}{{{D}_2} \Big(\frac{I}{A_o} e^{-t} \Big)^{-y}  } dt 
$.
Finally, resorting to the Laplace transform theory that \(F(s) = \mathcal{L}\left\{ e^{-t} f_1(t) \right\} = F_1(s+1)  \), where \(\mathcal{L}\left\{.\right\} \) denotes the Laplace operator and \(F_1(s) = (-1)^m \frac{\partial^{m}}{\partial s^{m}}f_1(t)  \), the closed-form expression for the PDF is provided in (\ref{j15}).
\section{Derivation of Outage Probability}
The outage probability for the considered TWR system can be obtained as
\begin{align*}
P_{\mathrm{out}} & =\text{Pr}\Big[\gamma_{\scaleto{\mathrm{S_1, T_2}}{3 pt}}< \gamma_{\scaleto{\mathrm{th}}{3 pt}}, \gamma_{\scaleto{\mathrm{S_2, T_2}}{3 pt}}<\gamma_{\scaleto{\mathrm{th}}{3 pt}} \Big]  = 1- 
\prod_{i=1}^{2} \Big(1-F_{\gamma_{\scaleto{\mathrm{S_i, T_2}}{5 pt}}}(\gamma_{\scaleto{\mathrm{th}}{3 pt}})\Big) \label{j7}
\addtag 
\end{align*}
Moreover, the statistics of \(\gamma_{\scaleto{\mathrm{S_1, T_2}}{3 pt}} \) can be written as
\begin{align*}
\text{Pr}\Big[\gamma_{\scaleto{\mathrm{S_1, T_2}}{3 pt}} & < \gamma \Big] = F_{\gamma_{\scaleto{\mathrm{S_1, T_2}}{3 pt}}} (\gamma) = 
\int_{0}^{\infty} F_{\gamma_{\scaleto{\mathrm{FSO}}{3 pt}}}\Bigg( \frac{(\kappa^{\mathrm{r}}+1)\gamma x }{\kappa^{\mathrm{r}}-x}\Bigg)  f_{\gamma_{\scaleto{\mathrm{I}}{3 pt}}}(x) dx \addtag \label{j2}
\end{align*}
Placing the requisites from (\ref{FSO-CDF}) and (\ref{intf-PDF}) into (\ref{j2}) and applying \cite[Eq. ((07.34.21.0088.01)]{Wolfram}, expression \(F_{\gamma_{\scaleto{\mathrm{S_1, T_2}}{3 pt}}} (\gamma)\) can be derived.  Similarly, the CDF of \(\gamma_{\scaleto{\mathrm{S_2, T_2}}{3 pt}} \) can be obtained as
\begin{align*}
\text{Pr}\Big[\gamma_{\scaleto{\mathrm{S_2, T_2}}{3 pt}} & < \gamma \Big] = F_{\gamma_{\scaleto{\mathrm{S_2, T_2}}{2 pt}}} (\gamma) = 
\int_{0}^{\infty} F_{\gamma_{\scaleto{\mathrm{RF}}{3 pt}}}\Bigg( \frac{(\gamma x)/|K_{\scaleto{\mathrm{1}}{3 pt}}^{\mathrm{t}}|^2 }{   (1- \frac{\gamma}{\kappa^{\mathrm{t}}}) } \Bigg)  f_{\gamma_{\scaleto{\mathrm{I}}{3 pt}}}(x) dx \addtag \label{j4}
\end{align*}
Placing the requisites from (\ref{RF-CDF}) and (\ref{intf-PDF}) into (\ref{j4}) and applying \cite[Eq. (07.34.21.0088.01)]{Wolfram}, the required closed-form can be derived. Plugging (\ref{j2}) and (\ref{j4}) into (\ref{j7}) and replacing \(\gamma \) by \(\gamma_{\scaleto{\mathrm{th}}{3 pt}} \), the closed-form expression for the outage probability can be derived as presented in (\ref{j23}).
\section{Derivation of ASR}
From (\ref{j1}), the closed-form expression for \(f_{\gamma_{\scaleto{\mathrm{S_1, T_2}}{5 pt}}}(x)\) can be obtained as follows
\begin{align*}
f_{\gamma_{\scaleto{\mathrm{S_1, T_2}}{5 pt}}}(x) \simeq  \frac{1+\kappa^{\mathrm{r}}}{\kappa^{\mathrm{r}}} \int_{0}^{\infty} z f_{\gamma_{\scaleto{\mathrm{FSO}}{3 pt}}}\left[\frac{x(1+\kappa^{\mathrm{r}})z}{\kappa^{\mathrm{r}}} \right] f_{ \gamma_{\scaleto{\mathrm{I}}{3 pt}}} (z) dz \addtag \label{a1}
\end{align*}
The PDF can be obtained by differentiating (\ref{FSO-CDF}) and substituting the resulting expression along with (\ref{intf-PDF}), the PDF \(f_{\gamma_{\scaleto{\mathrm{S_1, T_2}}{5 pt}}}(x)\) can be derived with the aid of \cite[Eq. (07.34.21.0088.01), (07.34.03.0001.01)]{Wolfram}. Placing the required terms from the derived PDF in  \(\mathcal{R}_{1} = \frac{1}{2 \text{log}(2)} \int_{0}^{\infty} \text{log}(1+\Delta x) f_{\gamma_{\scaleto{\mathrm{S_1, T_2}}{5 pt}}}(x) dx \), the \(\mathcal{R}_{1}\) can be derived using \cite[Eq. (07.34.21.0088.01)]{Wolfram}. Similarly, the expression of \(f_{\gamma_{\scaleto{\mathrm{S_2, T_2}}{5 pt}}}(x)\) required for calculating \(\mathcal{R}_{2}\), can be obtained as
\begin{align*}
f_{\gamma_{\scaleto{\mathrm{S_2, T_2}}{5 pt}}}(x) \simeq \frac{1}{|K_{\scaleto{\mathrm{1}}{3 pt}}^{\mathrm{t}}|^2} \int_{0}^{\infty} z f_{\gamma_{\scaleto{\mathrm{RF}}{3 pt}}}\left[\frac{xz}{|K_{\scaleto{\mathrm{1}}{3 pt}}^{\mathrm{t}}|^2} \right] f_{ \gamma_{\scaleto{\mathrm{I}}{3 pt}}} (z) dz
\addtag \label{a2}
\end{align*}
Again, substituting (\ref{RF}) and (\ref{intf-PDF}) into (\ref{a2}), and applying \cite[Eq. (07.34.21.0088.01)]{Wolfram}, the closed-form expression for \(f_{\gamma_{\scaleto{\mathrm{S_2, T_2}}{5 pt}}}(x)\) can be derived. Plugging the resulting expression into \(\mathcal{R}_{2} = \frac{1}{2 \text{log}(2)} \int_{0}^{\infty} \text{log}(1+ x) f_{\gamma_{\scaleto{\mathrm{S_2, T_2}}{5 pt}}}(x) dx \), the \(\mathcal{R}_{2}\) can be derived with the help of \cite[Eq. (07.34.21.0088.01)]{Wolfram}.

\bibliographystyle{IEEEtran}
\bibliography{ar}

\end{document}